\documentclass[reqno,dvips,12pt]{article} %% english should be theLAST
\usepackage{epsfig,graphicx,amsmath,amssymb,amsfonts}
cm \textwidth=16.0 true cm \emergencystretch=10pt

\newtheorem{theorem}{\qquad Theorem}
\newtheorem{lemma}{\qquad Lemma} %[section]

\newcommand{\cA}{{\cal A}}
\newcommand{\cM}{{\cal M}}

\newcommand{\cN}{{\cal N}}
\newcommand{\cP}{{\cal P}}

\newcommand{\const}{\mathop{\rm const}\nolimits}

\newcommand{\nn}{\nonumber}

\newcommand{\pa}{\partial}

\begin{document}

\title {\textbf{Dimension of Holes and High-Temparature  Condensate  in Bose-Einstein Statistics}}

\author{\textbf{V.P.Maslov}\thanks{Moscow State University,
Physics Department, v.p.maslov@mail.ru}}
\date{ }
%\emph{Moscow Institute of Electronics and Mathematics, pm@miem.edu.ru}

\maketitle
\begin{abstract}
We introduce the notion of weight for the lattice dimension and
the notion of  topological dimension -- hole dimension. The
condensate in  Bose--holes exists in the case when temperature is
not low.
\end{abstract}

In our  previous papers we applied  the quantum statistics
approach to linguistic statistics \cite{Lingvostat}.

To each word a frequency dictionary assigns its  number of
occurrences in the corresponding corpus of texts. There may be
several words with the same number of occurrences.

The rank of a word (i.e., the order number of a word in a
frequency dictionary) is measured from the word $A_1$ with the
highest number of occurrences $s$ in a given  corpus of texts. The
number of all words with the same number of occurrences $i$ we
designate as $N_i$. Then, if $A_m$ is the last word in a list of
words with number of occurrences $m$ (the words of the same
number of occurrences can be arranged in an arbitrary  way), then
the rank $r_m$ of $A_m$ is obviously
$$
r_m= \sum_{i=m}^s N_i.
$$

There is an analogy between the Bose particles at the energy
level of an oscillator $\lambda_i=i$ and the words with the
occurrence number~$i$, namely, the words with the same occurrence
number can be ordered in an arbitrary way, say, alphabetically,
inversely alphabetically, or in any other order. The indexing of
the ranks (the indices) of words within the family of a given
occurrence number is arbitrary. In this sense, the words are
indistinguishable and are distributed according to the Bose
statistic.

However, there is a difference between the approaches under
consideration. In the frequency dictionary one evaluates the
number of occurrences of every word and then orders the words,
beginning with the most frequently occurring words.

When there were no computers, it was difficult for a person to
evaluate the number of words with equal occurrence number. By
looking at a page as if it were a picture, a person can determine
a desired word on this page by its graphical form at every place
of occurrence of the word. In this case, the person looks at a
page of the text as if it were a photo, without going into the
meaning. Similarly, if a person looks for a definite name in a
long list of intrants who had entered a college, this person
finds (or does not find) the desired name by eyes rather than
reads all the names one after another.

An eye gets into the way of recognizing the desired image, and
this ability intensifies as the viewed material increases: the
more pages the eye scans, the less is the difficulty in finding
the desired graphical form. Therefore, under a manual counting,
it was simpler to recognize the desired word on a page without
reading the text and to cross it out by a pencil, simultaneously
counting the number of occurrences of the word. This procedure is
repeated for any subsequent word, already using the text with the
words crossed out (``holes''), which facilitates the search. In
other words, the procedure is in the recognition of the image of
the given word, similar to the recognition, say, of a desired
mushroom in a forest without sorting out all the plants on the
soil one after another. An ordinary computer solves problems of
this kind by exhaustion, whereas a quantum computer
(see~\cite{Belavkin_Mas,Bel_Mas_Arx}) makes this by recognizing
the image.

However, for an ordinary computer, the number of operations
needed to find the occurrence frequency of a word is less than
the number of operations needed to find the number of words in
the text with a given occurrence frequency.

One can say that the number of mushrooms we gathered (took away
from the forest) is the number of holes we left in the forest.
Similarly, the words we had ``got out'' from the text in the
above way is an analog of holes rather than particles. Therefore,
the linguists count the rank of words starting from the opposite
end as compared to the starting end which would be used by
physicists. The physicists would count the particles starting
from the lowest level, whereas the holes, the absent electrons,
would be counted from the highest level.

For this reason, the words in a frequency dictionary are
associated with holes rather than particles. Correspondingly, the
dimension in the distribution of frequency dictionaries is to be
chosen as a ``hole'' dimension (''Dirac's hole''), which is
negative.

We will prove a cumulative formula in which the densities coincide
in shape with the Bose--Einstein distribution. The difference
consists only in that, instead of the set $\lambda_n$ of random
variables or eigenvalues of the Hamiltonian operator, the
Bose--Einstein formula contains some of their averages over the
cells \cite{Landau1}. In view of the theorem given below,  one can
proof that  the $\varepsilon_i$, which are averages of the energy
$\lambda_k$ at the $i$th cell, are nonlinear averages in the
sense of Kolmogorov~\cite{NelinSred}.

As in \cite{NegDimen}, the values of the random variable
$\lambda_1, \dots, \lambda_s$ are ordered in absolute value. Some
of the numbers $\lambda_1,\dots,\lambda_s$ may coincide. Then
these numbers are combined adding the corresponding
''probabilities'', i.e., the ratio of the number of ``hits''
at~$\lambda_i$ to the general number of trials. The number of
equal $\lambda_i: \lambda_i=\lambda_{i+1}= \dots = \lambda_{i+k}$
will be called the multiplicity $q_i$ of the value $\lambda_i$.
In our consideration, both the number of trials $N$ and~$s$ tend
to infinity.

Let $N_i$ be the number of ''appearances'' of the value
$\lambda_i: \ \lambda_i < \lambda_{i+1}$, then
\begin{equation}
\sum^s_{i=1} \frac{N_i}{N} \lambda_i=M, \label{Zipf1}
\end{equation}
where $M$ is the mathematical expectation.

The cumulative probability $\cP_k$ is the sum of the first~$k$
probabilities in the sequence $\lambda_i$: $\cP_k=\frac 1N
\sum_{i=1}^k N_i$, where $k<s$. We denote $NP_k=B_k$.

If all the variants for which
\begin{equation}\label{A}
\sum_{i=1}^s N_i = N
\end{equation}
and
\begin{equation}\label{B}
\sum_{i=1}^s N_i \lambda_i \leq E, \ \ E=MN\leq N
\overline{\lambda},
\end{equation}
where $\overline{\lambda}=\frac{\sum_{i=1}^s q_i \lambda_i}{Q}$,
$Q=\sum_{i=1}^s q_i$, are equivalent (equiprobable), then
\cite{NelinSred,MatZamTheor,NoPredp} the majority of the variants
will accumulate near the following dependence of the ''cumulative
probability'' $B_l\{N_i\}=\sum_{i=1}^l N_i$,
\begin{equation}
\sum_{i=1}^l N_i= \sum_{i=1}^l
\frac{q_i}{e^{\beta'\lambda_i-\nu'}-1}, \label{Zipf2}
\end{equation}
where $\beta'$ and $\nu'$ are determined by the conditions
\begin{equation}\label{Zipf2a}
B_s=N,
\end{equation}
\begin{equation}\label{Zipf2a'}
\sum_{i=1}^s \frac{q_i \lambda_i}{e^{\beta' \lambda_i-\nu'}-1}=E,
\end{equation}
as $N \to \infty$ and $s \to \infty$.

We introduce the notation: $\cM$ is the set of all sets $\{N_i\}$
satisfying conditions~(\ref{A}) and~(\ref{B}); \ $\cN\{\cM\}$ is
the number of elements of the set~$\cM$.

\begin{theorem} \label{theor1}
Suppose that all the variants of sets $\{N_i\}$ satisfying the
conditions ~(\ref{A}) and ~(\ref{B}) are equiprobable. Then the
number of variants $\cN$ of sets $\{N_i\}$ satisfying
conditions~(\ref{A}) and~(\ref{B}) and the additional relation
\begin{equation} \label{theorema1}
|\sum^l_{i=1} N_i - \sum^l_1\frac{q_i}{e^{\beta'
\lambda_i-\nu'}-1}|\geq  N^{(3/4+\varepsilon)}
\end{equation}
is less than $\frac{c_1 \cN\{\cM\}}{N^m}$ (where~$c_1$ and~$m$
are any arbitrary numbers, $\sum_{i=1}^l q_i \geq\varepsilon Q$,
and $\varepsilon$ is arbitrarily small).
\end{theorem}

{\bf{\qquad Proof of Theorem 1.}}

Let $\cA$ be a subset of $\cM$ satisfying the condition
$$
|\sum_{i=l+1}^s N_i - \sum_{i=l+1}^s \frac{q_i}
{e^{\beta\lambda_i-\nu}-1}|\leq \Delta;
$$
$$
|\sum_{i=1}^l N_i-\sum_{i=1}^l \frac{q_i}
{e^{\beta'\lambda_i-\nu'}-1}|\leq \Delta,
$$
where $\Delta$, $\beta$, $\nu$ are some real numbers independent
of~$l$.

We denote
$$
|\sum_{i=l+1}^s N_i-\sum_{i=l+1}^s \frac{q_i}
{e^{\beta\lambda_i-\nu}-1}| =S_{s-l};
$$
$$
|\sum_{i=1}^l N_i-\sum_{i=1}^l \frac{q_i}
{e^{\beta'\lambda_i-\nu'}-1}| =S_l.
$$

Obviously, if $\{N_i\}$ is the set of all sets of integers on the
whole, then
\begin{equation}\label{Proof1}
\cN\{\cM \setminus \cA\} = \sum_{\{N_i\}} \Bigl(
\Theta(E-\sum_{i=1}^s N_i\lambda_i) \delta_{(\sum_{i=1}^s N_i),N}
\Theta(S_l-\Delta)\Theta (S_{s-l}-\Delta)\Bigr),
\end{equation}
where $N_i$ are  arbitrary integers.

Here the sum is taken over all integers $N_i$, $\Theta(\lambda)$
is the Heaviside function, and $\delta_{k_1,k_2}$ is the
Kronecker symbol.

We use the integral representations
\begin{eqnarray}
&&\delta_{NN'}=\frac{e^{-\nu N}}{2\pi}\int_{-\pi}^\pi d\varphi
e^{-iN\varphi} e^{\nu N'}e^{i N'\varphi},\label{D7}\\
&&\Theta(y)=\frac1{2\pi i}\int_{-\infty}^\infty
d\lambda\frac1{\lambda-i}e^{\beta y(1+i\lambda)}.\label{D8}
\end{eqnarray}

Now we perform the standard regularization. We replace the first
Heaviside function~$\Theta$ in~(\ref{Proof1}) by the continuous
function
\begin{equation}
\Theta_{\alpha}(y) =\left\{
\begin{array}{ccc}
0 &\mbox{for}& \alpha > 1, \  y<0 \nn \\
1-e^{\beta y(1-\alpha)} &\mbox{for}& \alpha > 1,\  y \geq 0,
\label{Naz1}
\end{array}\right.
\end{equation}
\begin{equation}
\Theta_{\alpha}(y) =\left\{
\begin{array}{ccc}
e^{\beta y(1-\alpha)} &\mbox{for}&\alpha < 0, \ y<0 \nn \\
1 &\mbox{for}& \alpha < 0, \ y \geq 0, \label{Naz2}
\end{array}\right.
\end{equation}
where $\alpha \in (-\infty,0) \cup (1, \infty)$ is a parameter,
and obtain
\begin{equation}\label{proof2}
\Theta_\alpha(y) = \frac1{2\pi i} \int_{-\infty}^{\infty}
e^{\beta y(1+ix)} (\frac 1{x-i} - \frac 1{x-\alpha i}) dx.
\end{equation}

If  $\alpha > 1$, then $\Theta(y)\leq \Theta_{\alpha}(y)$.

Let $\nu <0$. We substitute~(\ref{D7}) and~(\ref{D8})
into~(\ref{Proof1}),  interchange the integration and summation,
then pass to the limit as $\alpha \to \infty$ and obtain the
estimate
\begin{eqnarray}
&&\cN\{\cM \setminus   \cA\} \leq \nn \\
&&\leq \Bigl|\frac{e^{-\nu N+\beta E }}{i(2\pi)^2}\int_{-\pi}^\pi
\bigl[ \exp(-iN\varphi)
\sum_{\{N_j\}}\exp\bigl\{-\beta\sum_{j=1}^s
N_j\lambda_j+(i\varphi+\nu)
\sum_{j=1}^sN_j\bigr\}\bigr]\ d\varphi \times \nn \\
&& \times \Theta(S_l -\Delta)\Theta(S_{s-l}-\Delta)\Bigr|,
\end{eqnarray}
where $\beta$ and $\nu$ are real parameters such that the series
converges for them.

To estimate the expression in the right-hand side, we bring the
absolute value sign inside the integral sign and then inside the
sum sign, integrate over $\varphi$, and obtain
\begin{eqnarray}
&&\cN\{\cM \setminus \cA\} \leq \frac{e^{-\nu N+\beta E }}{2\pi}
\sum_{\{N_i\}}\exp\{-\beta\sum_{i=1}^sN_i\lambda_i+\nu
\sum_{i=1}^s N_i\}\times \nn \\
&& \times\Theta (S_l-\Delta)\Theta (S_{s-l}-\Delta).
\end{eqnarray}

We denote
\begin{equation}\label{D9a}
Z(\beta,N)=\sum_{\{N_i\}} e^{-\beta\sum_{i=1}^s N_i\lambda_i},
\end{equation}
where the sum is taken over all $N_i$ such that $\sum_{i=1}^s
N_i=N$,
$$
\zeta_l(\nu,\beta)= \prod_{i=1}^{l} \xi_i\left(\nu,\beta\right);
\zeta_{s-l}(\nu,\beta)= \prod_{i=l+1}^{s}
\xi_i\left(\nu,\beta\right);
$$
$$
\quad \xi_i(\nu,\beta)=
\frac{1}{(1-e^{\nu-\beta\lambda_i})^{q_i}}, \qquad i=1,\dots,l.
$$

It follows from the inequality for the hyperbolic cosine
$\cosh(x)=(e^x+e^{-x})/2$ for $|x_1| \geq \delta; |x_2| \geq
\delta$:
\begin{equation}
\cosh(x_1)\cosh(x_2)= \cosh(x_1+x_2) + \cosh(x_1 - x_2) >
\frac{e^\delta}{2} \label{D33}
\end{equation}
that the inequality
\begin{equation}
\Theta(S_{s-l}-\Delta) \Theta(S_{l}-\Delta)\le e^{-c\Delta}
\cosh\Bigl(c\sum_{i=1}^{l} N_i
-c\phi_l\Bigr)\cosh\Bigl(c\sum_{i=l+1}^{s} N_i
-c\overline{\phi}_{s-l}\Bigr), \label{D34}
\end{equation}
where
$$
\phi_l= \sum_{i=1}^l \frac{q_i}{e^{\beta'\lambda_i-\nu'}-1};
\qquad \overline{\phi}_{s-l}= \sum_{i=l+1}^s
\frac{q_i}{e^{\beta\lambda_i-\nu}-1},
$$
holds for all positive $c$ and~$\Delta$.

We obtain
\begin{eqnarray}
&&\cN\{\cM \setminus \cA\} \leq  e^{-c\Delta} \exp\left(\beta
E-\nu N\right) \times \nn \\
&& \times \sum_{\{N_i\}}\exp\{-\beta\sum_{i=1}^l
N_i\lambda_i+\nu\sum_{i=1}^l N_i\} \cosh\left(\sum_{i=1}^{l} c N_i -
c\phi\right) \times \nn \\
&& \times \exp\{-\beta\sum_{i=l+1}^s N_i\lambda_i +\nu
\sum_{i=l+1}^s
N_i\} \cosh\Bigl(\sum_{i=l+1}^s c N_i -c\overline{\phi}\Bigr) = \nn \\
&& =e^{\beta E} e^{-c\Delta} \times \nn \\
&&\times \left( \zeta_l(\nu-c,\beta) \exp(-c\phi_{l})
+\zeta_l(\nu+c,\beta)\exp(c\phi_{l})\right) \times \nn \\
&&\times\left(\zeta_{s-l}(\nu-c,\beta)
\exp(-c\overline{\phi}_{s-l})
+\zeta_{s-l}(\nu+c,\beta)\exp(c\overline{\phi}_{s-l})\right).
\label{5th}
\end{eqnarray}

Now we use the relations
\begin{equation}\label{5tha}
\frac {\pa}{\pa\nu}\ln \zeta_l|_{\beta=\beta',\nu=\nu'}\equiv
\phi_l; \frac {\pa}{\pa\nu}\ln
\zeta_{s-l}|_{\beta=\beta',\nu=\nu'}\equiv \overline{\phi}_{s-l}
\end{equation}
and the expansion $\zeta_l(\nu\pm c,\beta)$ by the Taylor formula.
There exists a $ \gamma <1$ such that
$$
\ln(\zeta_l(\nu\pm c,\beta)) =\ln\zeta_l(\nu,\beta)\pm
c(\ln\zeta_l)'_\nu(\nu,\beta)+\frac{c^2}{2}(\ln\zeta_l)^{''}_\nu
(\nu\pm\gamma c,\beta).
$$
We substitute this expansion, use formula~(\ref{5tha}), and see
that $\phi_{\nu,\beta}$ is cancelled.

Another representation of the Taylor formula implies
\begin{eqnarray}
&&\ln\left(\zeta_l(\nu+c,\beta)\right)=
\ln\left(\zeta_l(\beta,\nu)\right)+
\frac{c}\beta\frac{\pa}{\pa\nu}\ln\left(\zeta_l(\beta,\nu)\right)+\nn\\
&&+\int_{\nu}^{\nu+c/\beta}d\nu' (\nu+c/\beta-\nu')
\frac{\pa^2}{\pa\nu'^2}\ln\left(\zeta_l(\beta,\nu')\right).\label{CC1}
\end{eqnarray}
A similar expression holds for $\zeta_{s-l}$.

From the explicit form of the function $\zeta_l(\beta,\nu)$, we
obtain
\begin{equation}
\frac{\pa^2}{\pa\nu^2}\ln\left(\zeta_l(\beta,\nu)\right)=
\beta^2\sum_{i=1}^{l}
\frac{g_i\exp(-\beta(\lambda_i+\nu))}{(\exp(-\beta(\lambda_i+\nu))-1)^2}\leq
\beta^2Qd, \label{CC2}
\end{equation}
where $d$ is given by the formula
$$
d=\frac{\exp(-\beta(\lambda_1+\nu))}{(\exp(-\beta(\lambda_1+\nu))-1)^2}..
$$
The same estimate holds for $\zeta_{s-l}$.

Taking into account the fact that $\zeta_l\zeta_{s-l}=\zeta_s$,
we obtain the following estimate for $\beta=\beta'$ and
$\nu=\nu'$:
\begin{equation}\label{eval1}
\cN\{\cM \setminus \cA\}
\leq\zeta_s(\beta',\nu')\exp(-c\Delta+\frac{c^2}{2}\beta^2Qd)
\exp(E\beta'-\nu'N).
\end{equation}

Now we express $\zeta_s(\nu',\beta')$ in terms $Z(\beta,N)$. To
do this, we prove the following lemma.

\begin{lemma}%lemma 1
Under the above assumptions, the asymptotics of the integral
\begin{equation}\label{lemma_1}
Z(\beta,N) = \frac{e^{-\nu N}}{2\pi}\int_{-\pi}^\pi
 d\alpha e^{-iN\alpha}\zeta_s(\beta,\nu+i\alpha)
\end{equation}
has the form
\begin{equation}\label{lemma_2}
Z(\beta,N) = C e^{-\nu N} \frac{\zeta_s(\beta,\nu)}{|(\partial^2
\ln\zeta_s(\beta,\nu))/ (\partial^2\nu)|} (1+O(\frac 1N)),
\end{equation}
where $C$ is a constant.
\end{lemma}

The proof of the lemma readily follows from the saddle-point
method and the inequalities
\begin{equation}\label{nerav}
|\xi_i(\nu+i\alpha,\beta)| < \xi_i(\nu,\beta), \qquad
|\zeta_s(\nu+i\alpha,\beta)| <\zeta_s(\nu\,\beta),
\end{equation}
which hold, because $e^{\nu-\beta\lambda_i} <1$ for all
$\alpha\neq 2\pi n$, where $n$ is an integer. It follows from
these inequalities that $\alpha=0$ is a saddle point of
integral~(\ref{lemma_1})~\cite{Fedoruk,Oper_Meth}.

\begin{lemma}%lemma 2
The quantity
\begin{equation}\label{qqq5}
\frac{1}{\cN(\cM)} \sum_{\{N_i\}} e^{-\beta\sum_{i=1}^s
N_i\lambda_i},
\end{equation}
where $\sum N_i =N$ and $\lambda_iN_i\leq E-N^{1/2+\varepsilon}$,
tends to zero faster than $N^{-k}$ for any $k$, $\varepsilon>0$.
\end{lemma}

We consider the point of minimum in $\beta$ of the right-hand
side of ~(\ref{5th}) with $\nu(\beta,N)$ satisfying the condition
$$
\sum \frac{q_i}{e^{\beta\lambda_i-\nu(\beta,N)}-1} =N.
$$
It is easy to see that it satisfies condition~(\ref{Zipf2a}). Now
we assume that the assumption of the lemma is not satisfied.

Then for $\sum N_i=N$,  $\sum \lambda_i N_i\geq
E-N^{1/2+\varepsilon}$, we have
$$
e^{\beta E}\sum_{\{N_i\}} e^{-\beta\sum_{i=1}^s N_i\lambda_i}\geq
e^{(N^{1/2}+\varepsilon)\beta}.
$$
Obviously, $\beta\ll \frac{1}{\sqrt{N}}$ provides a minimum
of~(\ref{5th}) if the assumptions of Lemma~1 are satisfied, which
contradicts the assumption that the minimum in~$\beta$ of the
right-hand side of~(\ref{5th}) is equal to~$\beta'$.

We set $c=\frac\Delta{N^{1+\alpha}}$ in formula~(\ref{eval1})
after the substitution~(\ref{lemma_2}); then it is easy to see
that the ratio
$$
\frac{\cN(\cM \setminus\cA)}{\cN(\cM)}\approx \frac 1{N^m},
$$
where $m$ is an arbitrary integer, holds for
$\Delta=N^{3/4+\varepsilon}$. The proof of the theorem is
complete.

Now we consider the notion of the lattice dimension.

We consider a straight line, a plane, and a three-dimensional
space. We separate points $i=0,1,2, \dots$ on the line and points
$x=i=0,1,2,\dots$, \ $y=j=0,1,2, \dots$ on the coordinate axes
$x,y$ on the plane. We associate this set of points $(i,j)$ with
the points on the straight line (with the positive integers
$l=1,2 \dots$) up to the quantum constant~$\chi$ of the lattice.

According to M.~Gromov's definition~\cite{Gromov}, the asymptotic
(topological) dimension of this lattice is equal to two.

We associate each point with a pair of points~$i$ and~$j$
according to the rule $i+j=l$. The number of such points $n_l$ is
equal to $l+1$. In addition, we assume that $z=k=0,1,2,\dots$ on
the axis, i.e., we set $i+j+k=l$. In this case, the number of
points $q_l$ is equal to
$$
q_l =\frac{(l+1)(l+2)}{2}.
$$

If we set $\lambda_i=l$ in formula~(\ref{Zipf2}), then, in the
three-dimensional case, each~$i$ is associated with
$\frac{(l+1)(l+2)}{2}$ of mutually equal $x_l=l$ (these are the
multiplicities or the $q_l-$hold degeneracies of the spectrum of
the oscillator). Formula~(\ref{Zipf2}) in this special case
becomes
\begin{equation}\label{3.4}
N_l= const \sum_{i=0}^l \frac{(i+1)(i+2)}{2(e^{\beta i-\nu}-1)};
\end{equation}
\begin{equation}\label{3.5}
\Delta N_i= const \frac{(i+1)(i+2)}{2(e^{\beta
i-\nu}-1)}\Delta_i, \ \ \Delta_i=1,
\end{equation}
\begin{equation}\label{3.5a}
\Delta E_i=\const \frac{i(i+1)(i+2)}{2(e^{\beta i-\nu}-1)}\Delta_i
\end{equation}
for large~$i$, $\frac{\Delta_i}{i} \to 0$,
\begin{equation}\label{3.6}
dE= const \frac{\omega^3 d \omega}{e^{\beta\omega}-1}; \ \beta=
\frac hT
\end{equation}
(cf. formula~(60.4) in~\cite{Landau}).

Thus, we obtain a somewhat sharper version of the famous Planck
formula for the radiation of a black body.

For the $D$-dimensional case, it is easy to verify that the
sequence of weights (multiplicities) of the number of versions
$i= \sum_{k=1}^D m_k$, where $m_k$ are arbitrary positive
integers, has the form of the binomial coefficient
\begin{equation}\label{3.7}
q_i(D) = const\frac{(i+D-1)!}{i!D!},
\end{equation}
where the constant depends on~$D$.

Thus, for any $D$, formula~(\ref{Zipf2}) has the form
\begin{equation}\label{3.8}
N_l = const \sum_{i=1}^l  \frac{q_i (D)}{e^{\beta i}-1}.
\end{equation}

For the positive integers, we have a sequence of weights~$q_i$
(or, simply, a weight) of the form~(\ref{3.7}).

Our weight series can easily be continued to an arbitrary case by
replacing the factorials with the $\Gamma$-functions; in this
case, we assume that~$D$ is negative.

This is the negative topological dimension (the hole dimension)
of the quantized space (lattice).

If $D>1$, then, as $i\to\infty$, a condensation of a sufficiently
small perturbation occurs in the spectrum of the oscillator and
the multiplicities split, i.e., the spectrum becomes denser
as~$i$ increases. The fact that~$D$ is negative means that there
is strong rarefaction in the spectrum as $i\to\infty$ (the
constant in formula~(\ref{3.8}) must be sufficiently large).

For non-positive integer~$D$, the terms $i=0,1,2,3, \dots, -D$
become infinite. This means that they are very large in the
experiment, which permits determining the lattice negative
dimension corresponding to a given problem. We note that a new
condensate occurs, which is possible for small~$\beta$.

Now we return to frequency dictionaries. A frequency dictionary is
composed using texts from a certain array. Each word in the
dictionary is associated with the number of its encounters in the
original array of texts. We denote this number by~$\omega_i$ and
the number of words corresponding to this number of encounters
by~$N_i$. The sum of all~$N_i$ is equal to the volume of the
entire dictionary, and the sum of products $\omega_i n_i$ is
equal to the volume of the entire array of texts used to compose
the dictionary.

The number of words encountered only once in the array of texts
is approximately equal to 1/3  of the entire frequency dictionary
which the  number of words equal to $N$. So  as $N\to\infty$ this
is the condensate. It follows from the above that $D = -1$ for the
dictionary. Hence, for $\beta\ll 1$ and $\nu \sim 1$, we have
\begin{equation} \label{rang1}
N_l = const\sum_{i=2}^l \frac{1}{i(i-1)(e^{\beta i-\nu}-1)} \sim
const\int^\omega
\frac{d\omega}{\alpha\omega(\alpha\omega-1)(e^{\beta\alpha\omega-\nu}-1)},
\end{equation}
where $\omega=l$ and $\alpha$ is the scale constant.  If $\omega$
is  finite and  $\beta\ll 1$ the integral may be taken.

For the frequency of ''Japanese candles'' of 30 stocks in the
stock market (see~\cite{ManZam_T&V}), the dimension is equal
to~$0$.

Figures 1 and 2 show the rank-frequency curves for  Leo Tolstoy's
War and Peace. In Fig. 1, the low-frequency part of the
dictionary is approximated by formula ~(\ref{rang1}) as $\beta=0$
(words with frequencies
> 3; $r_0 = 43408.8$, $\alpha = 0.891995$, and $c_1 =
-189.321$). For the entire dictionary, Fig. 2 shows the deviation
of the theoretical data from the dictionary data (the difference
between the frequency given in the dictionary and the frequency
given by the formula) against rank. The deviation is seen not to
exceed 1.5 words if the rank is larger than 300.
\begin{figure}[h] %WarAndPeace1
%\begin{minipage}{0.5\linewidth}
%\centering\epsfig{figure= S0x0p0,width=\linewidth}
\centering\epsfig{figure=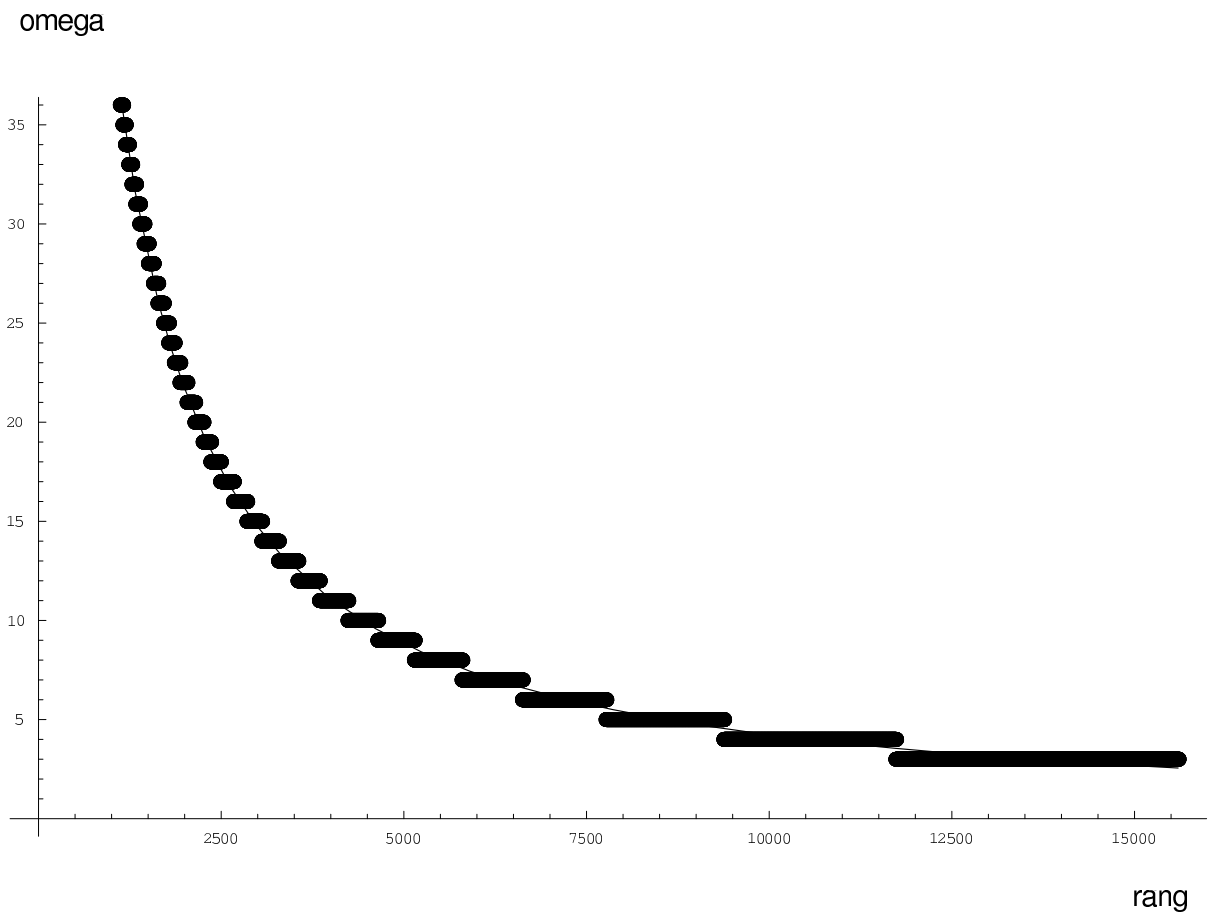,width=\linewidth}
\caption{\small War and Peace: the variable $r$ is the word's
number in the frequency dictionary in ascending order of
frequencies; $\omega$ is the word frequency}
%\end{minipage}\hfill
\end{figure}
\begin{figure}[h]%WarAndPeace2
%\begin{minipage}{0.5\linewidth}
%\centering\epsfig{figure= S0x0p0,width=\linewidth}
\centering\epsfig{figure=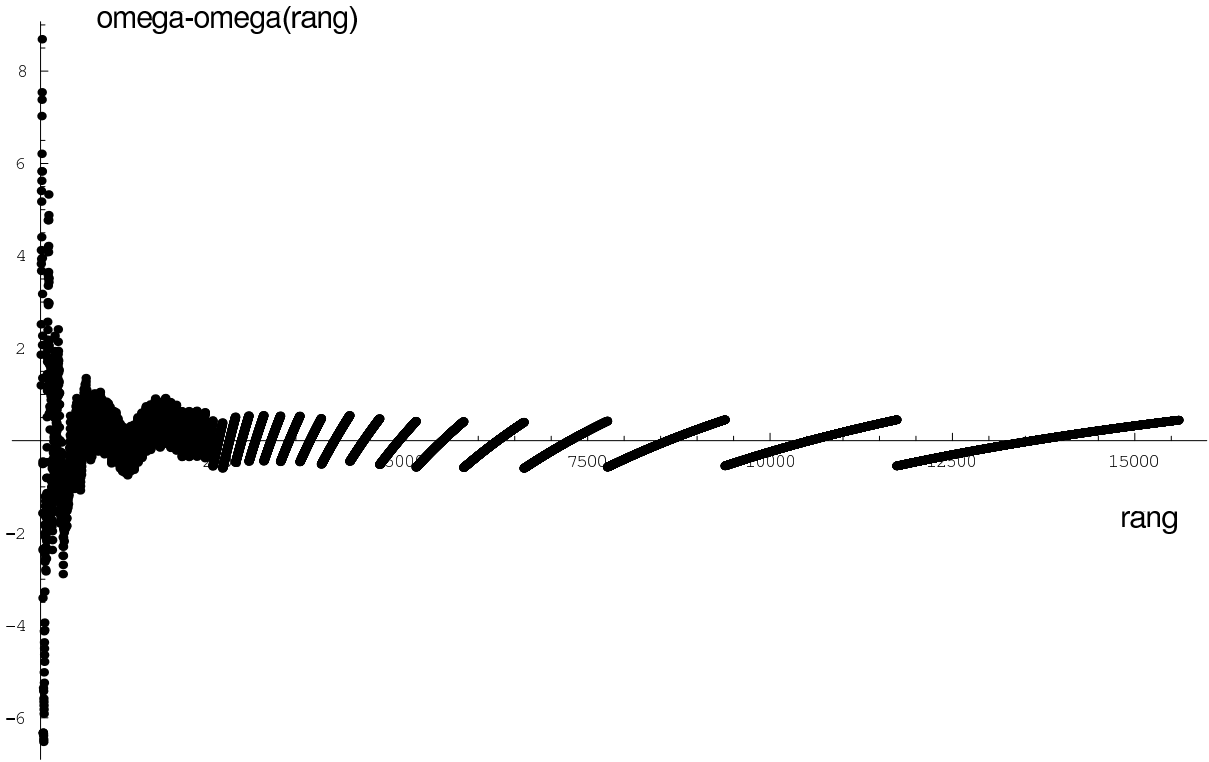,width=\linewidth}
\caption{\small The deviation of the theoretical data from the
dictionary data  for the first volume of War and Peace}
%\end{minipage}\hfill
\end{figure}

\begin{figure}[h] %''Japanese candles''
%\begin{minipage}{0.5\linewidth}
%\centering\epsfig{figure= S0x0p0,width=\linewidth}
\centering\epsfig{figure=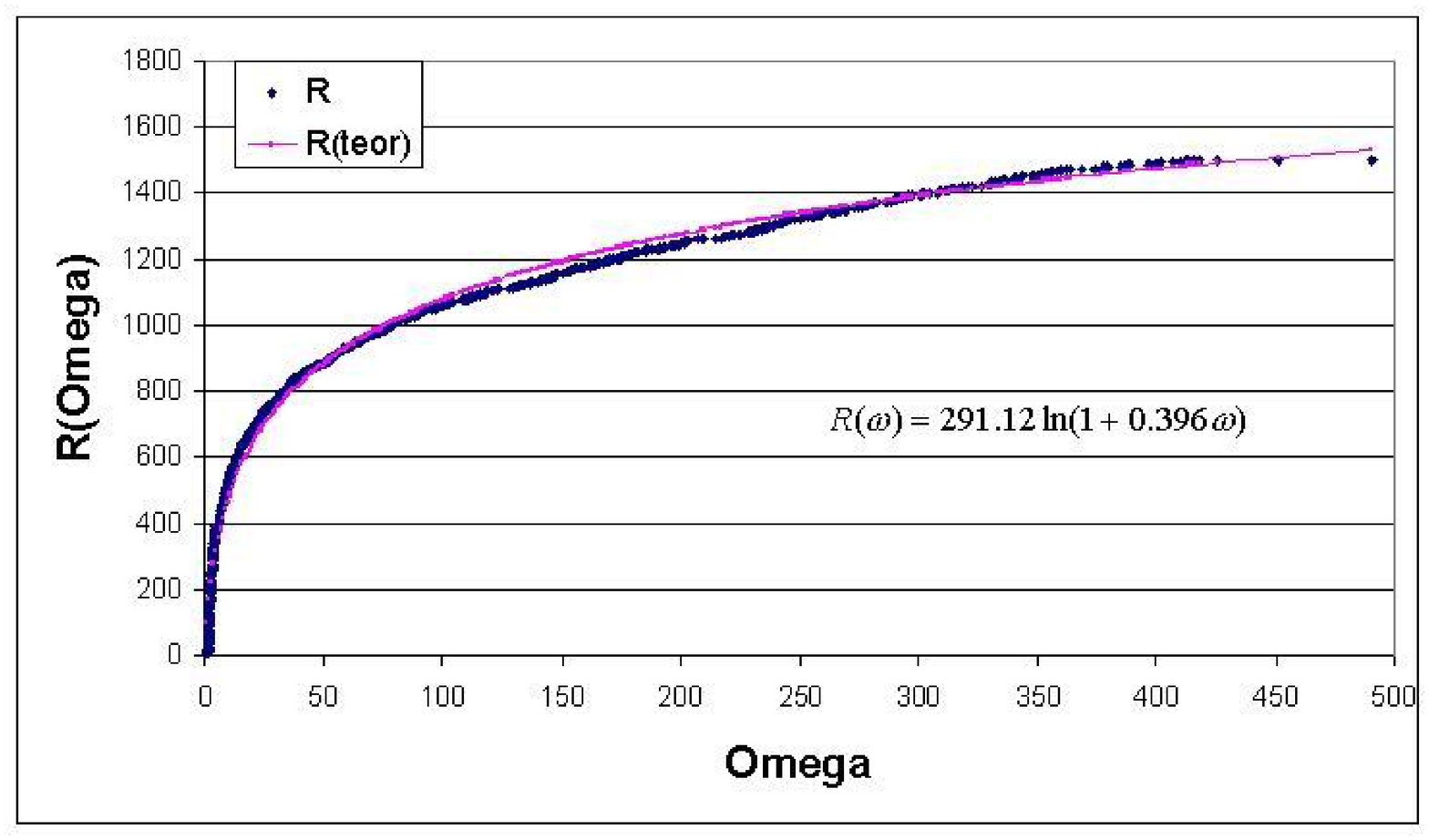 ,width=\linewidth}
\caption{\small The Japanese candle order number vs. frequency
for 30 stocks: $R$ is the candle order number, starting from the
smallest candle; $\omega$ is the candle frequency.}
%\end{minipage}\hfill
\end{figure}

\textbf{Examples of holes.}

In the Frenkel' theory of crystals, a hole (Dirac's hole) is an
absent electron, and it behaves as a particle moving along the
crystal. However, in contrast to an electron, this particle
cannot exist outside the crystal; there is no hole in vacuum. In
the same way, a hole, as we understand it, cannot exist without a
sufficiently large scale of spaces in which it is ``made.''

In fact, the fractal dimension of a hole, a crack in a rock, etc.,
depends on the scale at which ``our eye'' distinguishes points.
For instance, the coastline is of fractal nature when observed
from a plane. However, if we stand at the very coast, then this
line is smooth, and its dimension is equal to that of a smooth
curve, i.e., to one.

When considering cracks in metal, the dimension depends on the
resolving capacity of the device used to observe the
crack~\cite{Saouma2, Gold1, Gold2}. Denote the resolving capacity
of the device by $\kappa$. The usual definition of the Hausdorff
dimension is based on the condition that some balls (as a rule,
three-dimensional in practice) vanish. This means that the volume
$\Omega$ of the metal under consideration must be much greater
than~$\kappa^3$. Thus, reducing the problem to unit volume, we can
see that the radius  of the Hausdorff ball\footnote{That is, of
the ball used in the definition of the Hausdorff dimension,
see~\cite{Mandelbrot}.} ''tends to zero'' as~$\Omega\to\infty$.

The hole--wave propagation is especially visual when considering
elastic waves in media in which the Young modulus is sufficiently
large with respect to compression and vanishes with respect to
extension. In particular, sand has this property. The waves-holes
were studied in detail by the author together with P.~P.~Mosolov
in the case of one spatial coordinate (\cite{Masl_Mosol},
Appendix).

A worm corrodes a complicated serpiginous path-hole in the stem
of a tree. This path is very small with respect to the stem, and
we can assume that the thickness of the stem is infinitely large
as compared with the hole.

The dimension of a wire of the same thickness and the same
``sinuosity'' can be defined independently of the surrounding
medium, whereas one cannot consider the hole (made by the worm)
beyond the stem, although both the dimensions have the same
absolute value.

Let us now present an economical example.

Assume that somebody has inherited a great legacy in diverse forms
and frivols it away ``to the left and to the right.'' If it is
complicated to evaluate the amount of legacy, then the amount of
expenses increasing in time $t$ according to the power law $t^k$
(the appetite of expenses can increase, as well as the appetite
of profits!), then $k$ is the negative dimension, or $k$ is the
dimension of the ``hole'' thus occurring.

Let us now consider the simplest examples of  measures in the
general case for the $n$-dimensional space. Let $S_n$ be an
$n$-dimensional ball of radius~$r$. In the spherical coordinates,
the volume $\mu(S_n)$ of the ball is equal to $const \int_0^1
r^{n-1} dr= const \  r^n$. Here $r^{n-1}$ stands for the density.

In the sense of the Fourier transform, the multiplication by a
coordinate is dual to the corresponding derivation. Therefore, we
can speak of dual $n$ times differentiable functions in the
Sobolev space~$W_2^n$. Dirac distinguished between the left and
right components ''bra'' and ''ket'' in the ''bracket'' inner
product. The ``dual'' space of this space according to Dirac is
the space~$W_2^{-n}$ of Sobolev distributions (generalized
functions).

In the same way we can define the functions in $W_2^s$ by the
``inner product,'' where $s$ is a positive noninteger number, and
the space $W_2^{-s}$ as the ``inner product'' conjugate
to~$W_2^s$.

One can similarly proceed with the density (or the weight) $r^s$
and $r^{-s}$, by using, for instance, the Riesz kernel  to
represent functions in~$W_2^s$.

Let us present an example of a space (of noninteger positive
dimension) equipped with the Haar measure $r^\sigma$, where $0
\leq\sigma \leq 1$.

On the closed interval $0\leq x\leq 1$ there is a scale $0\leq
\sigma\leq 1$ of Cantor dust with the Haar measure equal to
$x^{\sigma}$ for any interval $(0,x)$ similar to the entire given
set of the Cantor dust. The direct product of this scale by the
Euclidean cube of dimension~$k-1$ gives the entire scale
$k+\sigma$, where $k \in \Bbb Z$ and~$\sigma \in (0,1)$. We
consider the space of negative dimension $-D=-k-\sigma$ with
respect to this very space. ``Quantizing'' of the densities
considered above leads to the density of the form~$\Gamma(D + l)/(\Gamma(D + 1)\Gamma(l + 1))$.

The following problem arises further: How one can understand a
scale of spaces in specific problems on a crack whose
dimension~$t$ increases as~$t\to\infty$? The size (volume) of the
metal or rock is not a dimension because all these objects are
three-dimensional. Where a sufficiently large value of dimension
arises?

For comparison, consider a gas in a three-dimensional vessel.
Every molecule of gas moves. The number of molecules is $\approx
10^{23}$. If a motion of every molecule is considered, then, as a
rule, one assumes that the entire dimension is equal to~$10^{69}$.

On the other hand, one can consider all molecules as points of
three-dimensional space. If we also take into account the
displacement at time~$t$, then we can consider the points in the
four-dimensional space as well. This very duality enabled us
(together with my co-author P.~P.~Mosolov) to pass from the
Newton equations for every molecule to equations of field-like
hydrodynamics of the entire gas~\cite{Mosol,Kvantovan_book}.

If we choose two moments of time for the displacement of the
molecules of gas in the $10^{69}$-dimensional space, i.e.,
consider the displacement of all $10^{23}$ molecules as a single
point at the first moment and as another point at the other
moment, then we can introduce the natural distance between these
points, i.e., a metric.

One can construct a scale of metrics between this metric and the
metric in three-dimensional space by combining diverse groups of
particles.

From the viewpoint of mechanics (elasticity theory) and the scales
typical for this science, metals, as well as rocks, consist of
``grains'' of sufficiently small (rather than atomic)
measure--volume. These grains can be of different size, and we
can regard the set of these grains as a point in the space of
sufficiently large dimension~\cite{Dauskardt, Saouma1}. This very
interpretation enables us to pass to the general definition of a
fractal crack as a hole and present the abstract mathematical
definition.

\textit{General definition of spaces of negative negative
dimension.} Let $M_{t_0}$ be a compactum, of Hausdorff
dimension~$t_0$, which is an element of a $t$-parameter scale of
mutually embedded compacta, $0< t < \infty$. Two scales of this
kind are said to be {\em equivalent\/} with respect to the
compactum $M_{t_0}$ if all compacta in these scales coincide for
any $t\geq t_0$. We say that the compactum $M_{t_0}$ is a {\em
hole\/} in this equivalent set of scales and the number $-t_0$ is
the {\em negative dimension\/} of this equivalence class.

\end{document}